\documentclass{rspublic}
\usepackage{amsmath,amsfonts,amssymb,latexsym,amsthm,graphicx}
\newcommand{\qudit}[1]{\left\vert #1 \right\rangle}

\newcommand{\Z}{\mathbb{Z}}

\newcommand{\C}{\mathbb{C}}
\newcommand{\I}{\mathbb{I}}

\begin{document}

\title[On a generalized SWAP gate]{On a generalized quantum {SWAP} gate}

\author[C.~M.~Wilmott]{Colin M.~Wilmott}

\affiliation{Institut f\"ur Theoretische Physik III,
Heinrich-Heine-Universit\"at, D\"usseldorf, Germany}

\label{firstpage}

\maketitle

\begin{abstract} {quantum computation; quantum circuits; quantum gates; modular binomial coefficients}
The {\small SWAP} gate plays a central role in network designs for
qubit quantum computation. However, there is a view to generalize
qubit quantum computing to higher dimensional quantum systems. In
this paper we construct a generalized {\small SWAP} gate using
only instances of the generalized controlled-{\small NOT} gate to
cyclically permute the states of $d$ qudits for $d$ prime.

\end{abstract}

\section{Introduction}

Of central importance to the theory of quantum computation is the
role assumed by multiple qubit gates in establishing a basis for
quantum network design. Moreover, the multiple qubit component
that best establishes itself as the hallmark  of quantum network
design  is the controlled-{\small NOT} ({\small CNOT}) gate. The
{\small CNOT} gate possesses a fundamental importance in the
theory of quantum computation assuming key roles in quantum
measurement and quantum error correction. Barenco \emph{et al.}
(1995) have shown that the {\small{CNOT}} gate is a principal
component in universal quantum gate constructions. Furthermore,
when we note that our ability to preserve quantum coherence rests
with our ability to successfully implement quantum computations,
the {\small{CNOT}} gate further distinguishes itself as the
hallmark multiple qubit gate as it is one of the few quantum gates
to have been experimentally realized within its coherence time
(Vatan \& Williams 2004).

A  standard feature of quantum computation  is to any express
multiple qubit gate in terms of single qubit gates and  the
{\small CNOT} gate (Barenco \emph{et al.} 1995). An example of
this is provided by the well-known {\small SWAP} gate which
describes the quantum operation that permutes the states of two
qubits. The {\small SWAP} gate  is seen as an important component
in the network design of Shor's algorithm (Fowler \emph{et al.}
2004), and  Liang \& Li (2005) maintain that successfully
implementing the {\small{SWAP}} gate is a necessary condition for
the networkability of quantum computation. Recently, it has been
asserted that there exist  advantages   in generalizing quantum
computation to higher dimension basis systems (Grassl \emph{et
al.} 2003). Therefore, considering new quantum network designs may
help to reveal the promise of qudit quantum computing. Indeed,
such new designs design may be merit in itself.

In this paper we concern ourselves with the design of a quantum
circuit to realize a generalized {\small SWAP} gate that
cyclically permutes $d$ qudit subsystems for $d$ prime. We
restrict ourselves to using only instances of the generalized
{\small CNOT} gate. Section \ref{prelim} introduces preliminary
material  as motivation for the design of quantum circuits which
exploit the generalized {\small CNOT} gate. Section \ref{WilNOT
section} introduces the design method for a quantum circuit to
realize a generalized {\small SWAP} of $d$ qudits for $d$ prime.
The analysis makes great use of modular binomial relationships to
achieve the desired result. Finally, \S\ref{wilnotd} revises the
design method of the previous section to achieve certain
permutations of $d$ qudits for $d$ other than prime.

\section{Preliminaries}\label{prelim}

\begin{figure}
\setlength{\unitlength}{0.08cm} \hspace*{65mm}
\begin{picture}(40,20)(23,10)
\put(0,20){\line(1,0){40}} \put(0,30){\line(1,0){40}}
\put(10,18){\line(0,1){11}} \put(10,20){\circle{4}}
\put(10,30){\circle*{2}} \put(20,20){\line(0,1){12}}
\put(20,20){\circle*{2}} \put(20,30){\circle{4}}
\put(30,18){\line(0,1){11}} \put(30,20){\circle{4}}
\put(30,30){\circle*{2}} \put(-7,29){$\qudit{\psi}$}
\put(-7,19){$\qudit{\phi}$} \put(44,29){$\qudit{\phi}$}
\put(44,19){$\qudit{\psi}$}
\end{picture}
\caption{The  {\small{SWAP}} gate illustrating the permutation of
two qubits through the use of three {\small CNOT} gates.  The
system begins in the state $\qudit{\psi}\otimes \qudit{\phi}$ and
ends in the state $\qudit{\phi}\otimes \qudit{\psi}$.}\label{swap}
\end{figure}
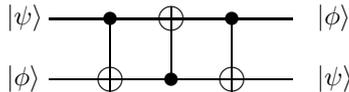

Let $\mathcal H$ denote the $d$-dimensional complex Hilbert space
$\C^d$. We fix each orthonormal basis state of the $d$-dimensional
space to correspond to an element of ring $\Z_d$ of integers
modulo $d$.  The basis $\{\qudit{0},\qudit{1},\dots,\qudit{d-1}\}
\subset \C^d$ whose elements correspond to the column vectors of
the identity matrix ${\I}_d$ is called the computational basis. A
\emph{qudit} is a $d$-dimensional quantum state $\qudit{\psi} \in
{\cal H}$ written as  $\qudit{\psi} =
\sum^{d-1}_{i=0}{\alpha_i\qudit{i}}$ where $\alpha_i \in \C$ and
$\sum^{d-1}_{i=0}{\vert\alpha_i\vert^2} = 1$. For a pair of qudits
$\qudit{\psi}, \qudit{\phi} \in {\cal H}$, the generalized {\small
CNOT} gate is a two-qudit quantum gate that acts on the state $
\qudit{\psi}\otimes \qudit{\phi}\in {\cal H} \otimes{\cal H}$. The
generalized {\small CNOT} gate  has control qudit $\qudit{\psi}$
and target qudit $\qudit{\phi}$ and its action on the basis states
$\qudit{m}\otimes\qudit{n} \in {\cal H} \otimes {\cal H}$ is given
by
\begin{eqnarray}
{\small \rm CNOT}\qudit{m}\otimes\qudit{n} =
\qudit{m}\otimes\qudit{n \oplus m}, \qquad m,n\in \Z_d,
\end{eqnarray}
with $\oplus$ denoting addition  modulo $d$. Figure~\ref{swap}
illustrates the role played by the {\small CNOT} gate in
describing the well-known {\small SWAP} gate for qubits.

We now introduce a generalized {\small SWAP} gate that cyclically
permutes the states of $d$ qudit subsystems for $d$ prime. The
construction process is restricted to using only instances of the
generalized {\small{CNOT}} gate.

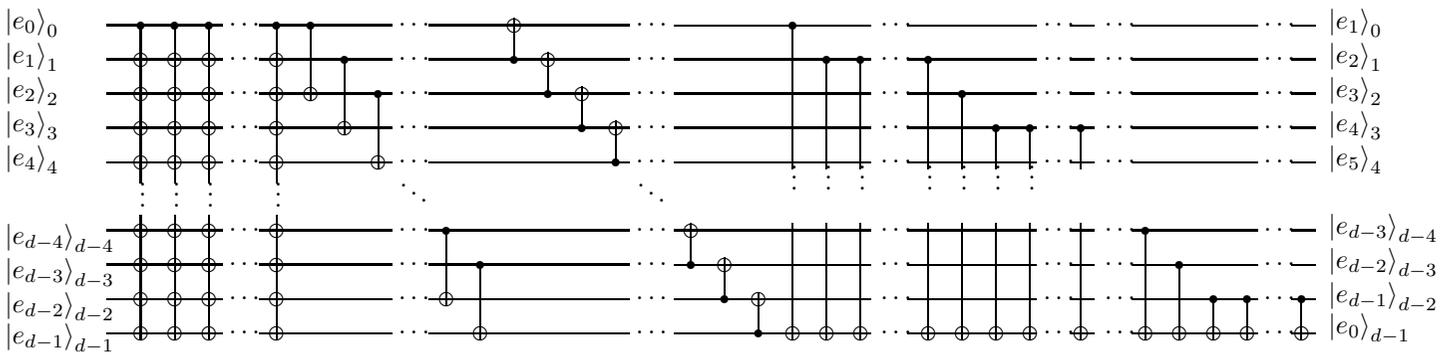
\begin{figure}
{ \setlength{\parindent}{1in} \setlength{\parskip}{40pt}

. \rotatebox{90}{\setlength{\unitlength}{0.045cm}
\hspace*{6mm}\begin{picture}(120,120)(285,45)
\put(95,120){\line(1,0){59}} \put(95,110){\line(1,0){59}}
\put(95,100){\line(1,0){59}} \put(95,90){\line(1,0){59}}
\put(95,60){\line(1,0){59}} \put(95,50){\line(1,0){59}}
\put(95,40){\line(1,0){59}} \put(95,30){\line(1,0){59}}
\put(95,80){\line(1,0){59}}
\put(120,110){\line(0,1){12}} \put(130,100){\line(0,1){12}}
\put(140,90){\line(0,1){12}} \put(150,80){\line(0,1){12}}
\put(156,120){\dots} \put(156,110){\dots} \put(156,100){\dots}
\put(156,90){\dots} \put(156,80){\dots} \put(156,30){\dots}
\put(156,40){\dots} \put(156,50){\dots} \put(156,60){\dots}
\put(120,110){\circle*{2}} \put(120,120){\circle{4}}
\put(130,100){\circle*{2}} \put(130,110){\circle{4}}
\put(140,90){\circle*{2}} \put(140,100){\circle{4}}
\put(150,80){\circle*{2}}
\put(150,90){\circle{4}}
%
\put(45,120){\line(1,0){39}} \put(86,120){\dots}
\put(86,110){\dots} \put(86,100){\dots} \put(86,90){\dots}
\put(86,80){\dots} \put(86,30){\dots} \put(86,40){\dots}
\put(86,50){\dots} \put(86,60){\dots}\put(0,120){\line(1,0){34}}
\put(0,110){\line(1,0){34}} \put(45,110){\line(1,0){39}}
\put(0,100){\line(1,0){34}} \put(45,100){\line(1,0){39}}
\put(0,90){\line(1,0){34}} \put(45,90){\line(1,0){39}}
\put(0,80){\line(1,0){34}} \put(45,80){\line(1,0){39}}
\put(9.5,67){\vdots} \put(19.5,67){\vdots} \put(29.5,67){\vdots}
\put(49.5,67){\vdots} \put(36,120){\dots} \put(36,110){\dots}
\put(36,100){\dots} \put(36,90){\dots} \put(36,80){\dots}
\put(36,30){\dots} \put(36,40){\dots} \put(36,50){\dots}
\put(36,60){\dots}
\put(0,30){\line(1,0){34}} \put(45,30){\line(1,0){39}}
\put(0,40){\line(1,0){34}} \put(45,40){\line(1,0){39}}
\put(0,50){\line(1,0){34}} \put(45,50){\line(1,0){39}}
\put(0,60){\line(1,0){34}} \put(45,60){\line(1,0){39}}
\put(10,30){\circle{4}} \put(10,120){\circle*{2}}
\put(10,110){\circle{4}} \put(10,100){\circle{4}}
\put(10,90){\circle{4}} \put(10,80){\circle{4}}
\put(10,40){\circle{4}} \put(10,50){\circle{4}}
\put(10,60){\circle{4}} \put(20,30){\circle{4}}
\put(20,120){\circle*{2}} \put(20,110){\circle{4}}
\put(20,100){\circle{4}} \put(20,90){\circle{4}}
\put(20,80){\circle{4}} \put(20,40){\circle{4}}
\put(20,50){\circle{4}} \put(20,60){\circle{4}}
\put(30,30){\circle{4}} \put(30,120){\circle*{2}}
\put(30,110){\circle{4}} \put(30,100){\circle{4}}
\put(30,90){\circle{4}} \put(30,80){\circle{4}}
\put(30,40){\circle{4}} \put(30,50){\circle{4}}
\put(30,60){\circle{4}} \put(50,30){\circle{4}}
\put(50,120){\circle*{2}} \put(50,110){\circle{4}}
\put(50,100){\circle{4}} \put(50,90){\circle{4}}
\put(50,80){\circle{4}} \put(50,40){\circle{4}}
\put(50,50){\circle{4}} \put(50,60){\circle{4}}
\put(60,120){\circle*{2}} \put(60,100){\circle{4}}
\put(60,98){\line(0,1){22}} \put(70,110){\circle*{2}}
\put(70,90){\circle{4}} \put(70,88){\line(0,1){22}}
\put(80,100){\circle*{2}} \put(80,80){\circle{4}}
\put(80,78){\line(0,1){22}} \put(85.9,67.5){$\ddots$}
\put(155.9,67.5){$\ddots$} \put(110,50){\circle*{2}}
\put(110,30){\circle{4}} \put(110,28){\line(0,1){22}}
\put(100,60){\circle*{2}} \put(100,40){\circle{4}}
\put(100,38){\line(0,1){22}} \put(10,28){\line(0,1){36.5}}
\put(20,28){\line(0,1){36.5}} \put(30,28){\line(0,1){36.5}}
\put(10,74){\line(0,1){47}} \put(50,28){\line(0,1){36.5}}
\put(50,74){\line(0,1){47}} \put(20,74){\line(0,1){47}}
\put(30,74){\line(0,1){47}}
\put(-30,119){$\qudit{e_0}_0$} \put(-30,109){$\qudit{e_1}_1$}
\put(-30,99){$\qudit{e_2}_2$} \put(-30,89){$\qudit{e_3}_3$}
\put(-30,79){$\qudit{e_4}_4$} \put(-30,27){$\qudit{e_{\small
d-1}}_{\small d-1}$} \put(-30,36){$\qudit{e_{\small d-2}}_{d-2}$}
\put(-30,56){$\qudit{e_{\small d-4}}_{d-4}$}
\put(-30,46){$\qudit{e_{\small d-3}}_{d-3}$}
\end{picture}}

.
 \rotatebox{90}{\setlength{\unitlength}{0.045cm} \hspace*{6mm}\begin{picture}(250,120)(-15,45)
\put(165,120){\line(1,0){29}} \put(165,110){\line(1,0){29}}
\put(165,100){\line(1,0){29}} \put(165,90){\line(1,0){29}}
\put(165,60){\line(1,0){29}} \put(165,50){\line(1,0){29}}
\put(165,40){\line(1,0){29}} \put(165,30){\line(1,0){29}}
\put(165,80){\line(1,0){29}} \put(194,120){\line(1,0){29}}
\put(194,110){\line(1,0){29}} \put(194,100){\line(1,0){29}}
\put(194,90){\line(1,0){29}}
\put(194,60){\line(1,0){29}} \put(194,50){\line(1,0){29}}
\put(194,40){\line(1,0){29}} \put(194,30){\line(1,0){29}}
\put(194,80){\line(1,0){29}} \put(226,30){\dots}
\put(226,40){\dots} \put(226,50){\dots} \put(226,120){\dots}
\put(226,80){\dots} \put(226,110){\dots} \put(226,100){\dots}
\put(226,90){\dots} \put(226,60){\dots}
\put(234,120){\line(1,0){38}} \put(234,110){\line(1,0){38}}
\put(234,100){\line(1,0){38}} \put(234,90){\line(1,0){38}}
\put(234,60){\line(1,0){38}} \put(234,50){\line(1,0){38}}
\put(234,40){\line(1,0){38}} \put(234,30){\line(1,0){38}}
\put(234,80){\line(1,0){38}} \put(200,28){\line(0,1){34}}
\put(210,28){\line(0,1){34}} \put(220,28){\line(0,1){34}}
\put(200,78){\line(0,1){42}} \put(210,78){\line(0,1){32}}
\put(220,78){\line(0,1){32}} \put(199.5,72){\vdots}
\put(209.5,72){\vdots} \put(219.5,72){\vdots}
\put(200,120){\circle*{2}} \put(200,30){\circle{4}}
\put(270,90){\circle*{2}} \put(270,30){\circle{4}}
\put(210,110){\circle*{2}} \put(210,30){\circle{4}}
\put(220,110){\circle*{2}} \put(220,30){\circle{4}}
\put(240,28){\line(0,1){34}} \put(250,28){\line(0,1){34}}
\put(260,28){\line(0,1){34}} \put(240,78){\line(0,1){32}}
\put(250,78){\line(0,1){22}} \put(260,78){\line(0,1){12}}
\put(239.5,72){\vdots} \put(249.5,72){\vdots}
\put(259.5,72){\vdots} \put(269.5,72){\vdots}
\put(240,110){\circle*{2}} \put(240,30){\circle{4}}
\put(250,100){\circle*{2}} \put(250,30){\circle{4}}
\put(260,90){\circle*{2}} \put(260,30){\circle{4}}
\put(190,30){\line(0,1){12}} \put(180,40){\line(0,1){12}}
\put(170,50){\line(0,1){12}} \put(170,50){\circle*{2}}
\put(170,60){\circle{4}} \put(180,40){\circle*{2}}
\put(180,50){\circle{4}} \put(190,30){\circle*{2}}
\put(190,40){\circle{4}} \put(270,28){\line(0,1){34}}
\put(270,78){\line(0,1){12}}
\put(285,30){\circle{4}} \put(285,90){\circle*{2}}
\put(285,28){\line(0,1){34}} \put(285,78){\line(0,1){12}}
\put(274,120){\dots} \put(274,110){\dots} \put(274,100){\dots}
\put(274,90){\dots} \put(274,80){\dots} \put(274,30){\dots}
\put(274,40){\dots} \put(274,50){\dots} \put(274,60){\dots}
\put(282,120){\line(1,0){7}} \put(282,110){\line(1,0){7}}
\put(282,100){\line(1,0){7}} \put(282,90){\line(1,0){7}}
\put(282,60){\line(1,0){7}} \put(282,50){\line(1,0){7}}
\put(282,40){\line(1,0){7}} \put(282,30){\line(1,0){7}}
\put(282,80){\line(1,0){7}} \put(292,120){\dots}
\put(292,110){\dots} \put(292,100){\dots} \put(292,90){\dots}
\put(292,80){\dots} \put(292,30){\dots} \put(292,40){\dots}
\put(292,50){\dots} \put(292,60){\dots}
\put(300,120){\line(1,0){37}} \put(300,110){\line(1,0){37}}
\put(300,100){\line(1,0){37}} \put(300,90){\line(1,0){37}}
\put(300,60){\line(1,0){37}} \put(300,50){\line(1,0){37}}
\put(300,40){\line(1,0){37}} \put(300,30){\line(1,0){37}}
\put(300,80){\line(1,0){37}} \put(304,60){\circle*{2}}
\put(304,30){\circle{4}} \put(314,50){\circle*{2}}
\put(314,30){\circle{4}} \put(324,40){\circle*{2}}
\put(324,30){\circle{4}} \put(334,40){\circle*{2}}
\put(334,30){\circle{4}} \put(304,28){\line(0,1){32}}
\put(314,28){\line(0,1){22}} \put(324,28){\line(0,1){12}}
\put(334,28){\line(0,1){12}} \put(339,120){\dots}
\put(339,110){\dots} \put(339,100){\dots} \put(339,90){\dots}
\put(339,80){\dots} \put(339,30){\dots} \put(339,40){\dots}
\put(339,50){\dots} \put(339,60){\dots}
\put(347,120){\line(1,0){7}} \put(347,110){\line(1,0){7}}
\put(347,100){\line(1,0){7}} \put(347,90){\line(1,0){7}}
\put(347,60){\line(1,0){7}} \put(347,50){\line(1,0){7}}
\put(347,40){\line(1,0){7}} \put(347,30){\line(1,0){7}}
\put(347,80){\line(1,0){7}} \put(350,28){\line(0,1){12}}
\put(350,40){\circle*{2}} \put(350,30){\circle{4}}
\put(358,119){$\qudit{e_1}_0$} \put(358,109){$\qudit{e_2}_1$}
\put(358,99){$\qudit{e_3}_2$} \put(358,89){$\qudit{e_4}_3$}
\put(358,79){$\qudit{e_5}_4$} \put(358,30){$\qudit{e_0}_{d-1}$}
\put(358,39){$\qudit{e_{d-1}}_{d-2}$}
\put(358,59){$\qudit{e_{d-3}}_{d-4}$}
\put(358,49){$\qudit{e_{d-2}}_{d-3}$}
\end{picture}}
} \caption{A generalized {\small SWAP gate} composed entirely in
terms of the generalized {\small CNOT gate} that cyclically
permutes the states of  $d$ qudit subsystems. }\label{wilnotswap}
\end{figure}

\section{A generalized {SWAP} gate }\label{WilNOT section}

We  construct  a generalized quantum {\small SWAP} gate that
cyclically permutes the states of $d$ qudit subsystems. Figure
\ref{wilnotswap} illustrates a generalized quantum {\small SWAP}
gate of $d$ qudit subsystems. The design method is restricted to
using only generalized {\small CNOT} gates. We suppose that the
first quantum system ${\mathcal{A}}_0$ prepared in the state
$\qudit{e_0}_0$, the second system ${\mathcal{A}}_1$ prepared in
the state $\qudit{e_1}_1$ and so forth, with the final system
${\mathcal{A}}_{d-1}$ prepared in the state
$\qudit{e_{d-1}}_{d-1}$. The process describes a quantum network
that uses only generalized {\small CNOT} gates to realize a
generalized {\small SWAP} of $d$ qudits for $d$ prime with the
result that the system ${\mathcal{A}}_0$ is in the state
$\qudit{e_1}_0$, the system ${\mathcal{A}}_1$ is in the state
$\qudit{e_2}_1$ and so forth, until the system
${\mathcal{A}}_{d-1}$ is in the state $\qudit{e_0}_{d-1}$. We make
use of the following result.

\begin{lemma} (Rosen \& Michaels 2000) $\sum^{k}_{n=0}{{l+n\choose n}} = {l+k+1\choose k}$.
\end{lemma}
\begin{theorem}\label{WilNOT}
Let $d = p$ be a prime. We provide an algorithm for the
construction of a generalized {\small SWAP} gate using only
instances of the generalized {\small CNOT} gate. The generalized
{\small SWAP} gate has

Input: $\qudit{e_k}_k; \ k = 0,\dots,d-1$

Output: $\qudit{e_{k+1}}_k; \ k = 0,\dots,d-2, \  \qudit{e_0}_{d-1}.$\\

The generalized {\small SWAP} gate algorithm is described as follows:\\

Input: $e_k := i_k^0; \  k = 0,\dots,d-1$

Output: $i_k^{d+2} = e_{k+1}; \  k = 0,\dots,d-1, \
i_{d-1}^{d+2} = e_0.$\\

\begin{description}
\item[Stage 1]{}\underline{Initialization} $j$ = 0.
\begin{eqnarray} e_k &:=&i^0_{k}\nonumber
\end{eqnarray}
for $k = 0, \dots, d-1.$ The algorithm initiates at stage 1, step
$j$ = 0 by  making the correspondence between a representative
input element $i_k^0$ of the algorithm and each standard basis
state $e_{k}$.

\item[Stage 2] $j = 1, \dots, d-1$.
\begin{eqnarray}
i^{j}_{0} &=& i^{j-1}_{0}\nonumber\\
i^{j}_{k} &=& i^{j}_{k-1}+i^{j-1}_{k}; \ k = 1,\dots d-1.\nonumber
\end{eqnarray}

Stage 2 consists of $d-1$  steps which repeat the sequence of
gates of step $j = 1$.  The sequence of gates at step $j = 1$, see
figure \ref{2.1}, is targeted on systems
${\mathcal{A}}_1,\dots,{\mathcal{A}}_{d-1}$. Each step of figure
\ref{2.1} is a composition of generalized {\small CNOT} gates
acting on consecutive pairs of systems and is written as a
shorthand form to represent a sequence of generalized {\small
CNOT} gates as illustrated in figure \ref{qwerty}. The algorithm
process of step $j = 1$ transforms the input sequence
$i^0_0,i^0_1,i^0_2,\dots, i_{d-1}^0$ to the state given by
$i_0^0,\sum_{k=0}^{1}{}i_k^0, \sum_{k=0}^{2}{}i_k^0, \dots,
\sum_{k=0}^{{d}-1}{}i_k^0$. Similarly,  the algorithm at step $j =
2$ takes the output from step $j = 1$ as input and repeats the
sequence of gates. The resulting state of the circuit at step $j =
2$ is given by $i_0^0$, $i_0^0+\sum_{k=0}^{1}{}i_k^0$,
$i_0^0+\sum_{k=0}^{1}{}i_k^0+\sum_{k=0}^{2}{}i_k^0$, $\dots,
i_0^0+\sum_{k=0}^{1}{}i_k^0+\sum_{k=0}^{2}{}i_k^0+\dots+\sum_{k=0}^{{d}-1}{}i_k^0$.
This process continues to step $j = d-1$. Figure \ref{stage2}
illustrates  initialization  on the circuit and the subsequent
$d-1$ steps of stage 2.

\item[Stage 3] $j = d$.
\begin{eqnarray}
i^{d}_{0} &=& i^{d-1}_{0}\nonumber\\
i^{d}_{1} &=& i^{d-1}_{1}\nonumber\\
i^{d}_{k} &=& i^{d}_{k-2} + i^{d-1}_{k}; \ k =
2,\dots,d-1.\nonumber
\end{eqnarray}

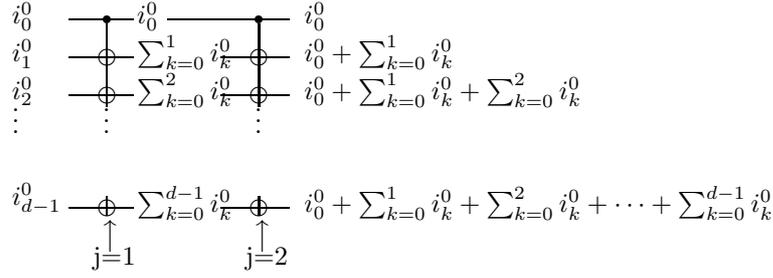
\begin{figure}\setlength{\unitlength}{0.05cm}
\hspace*{5mm}
\begin{picture}(60,90)(-30,-45)

\put(0,30){\line(1,0){17}}\put(0,20){\line(1,0){17}}
\put(0,-10){\line(1,0){17}}\put(0,40){\line(1,0){17}}

\put(40,30){\line(1,0){18}}\put(40,20){\line(1,0){18}}
\put(40,-10){\line(1,0){18}}\put(26,40){\line(1,0){32}}

\put(10,17){\line(0,1){22}} \put(10,20){\circle{4}}
\put(50,20){\circle{4}}\put(50,30){\circle{4}}\put(50,-10){\circle{4}}
\put(10,40){\circle*{2}} \put(50,40){\circle*{2}}
\put(50,17){\line(0,1){22}}
\put(10,30){\circle{4}}
\put(18,39){${i_0^0}$} \put(18,29){$\sum^{1}_{k=0}{i_k^0}$}
\put(18,19){$\sum^{2}_{k=0}{i_k^0}$}
\put(18,-11){$\sum^{d-1}_{k=0}{i_k^0}$}

\put(62,39){${i_0^0}$} \put(62,29){$i_0^0+\sum^{1}_{k=0}{i_k^0}$}
\put(62,19){$i_0^0+\sum^{1}_{k=0}{i_k^0}+\sum^{2}_{k=0}{i_k^0}$}
\put(62,-11){$i_0^0+\sum^{1}_{k=0}{i_k^0}+\sum^{2}_{k=0}{i_k^0}+\dots
+\sum^{d-1}_{k=0}{i_k^0}$}

\put(10,-10){\circle{4}}
\put(10,-12){\line(0,1){5}}
\put(50,-12){\line(0,1){5}}
\put(-15,29){$i_1^0$} \put(-15,19){$i_2^0$}
\put(-15,39){$i_0^0$}\put(-15,10){\vdots}
\put(9,10){\vdots}\put(49,10){\vdots} \put(-15,-9){$i_{d-1}^0$}
\put(8.5,-19){$\left\uparrow\begin{matrix} \vspace*{.1mm}
\cr\end{matrix}\right.$}\put(48.5,-19){$\left\uparrow\begin{matrix}
\vspace*{.1mm} \cr\end{matrix}\right.$
}\put(6.5,-25){j=1}\put(46.5,-25){j=2}
\end{picture}
\vskip-2em\caption{Generalized {\small SWAP} gate; stage 2, steps
$j = 1, 2$.}\label{2.1}
\end{figure}

\begin{figure}\setlength{\unitlength}{0.05cm}
\hspace*{2mm}
\begin{picture}(60,90)(-70,-40)

\put(0,30){\line(1,0){17}} \put(0,20){\line(1,0){17}}
\put(0,-10){\line(1,0){17}} \put(0,40){\line(1,0){17}}

\put(40,30){\line(1,0){48}} \put(40,20){\line(1,0){48}}
\put(40,-10){\line(1,0){48}} \put(40,40){\line(1,0){48}}

\put(10,17){\line(0,1){22}} \put(10,20){\circle{4}}
\put(60,20){\circle{4}} \put(50,30){\circle{4}}

\put(10,40){\circle*{2}} \put(50,40){\circle*{2}}
\put(60,30){\circle*{2}} \put(70,20){\circle*{2}}
\put(50,28){\line(0,1){12}} \put(60,18){\line(0,1){12}}
\put(70,17){\line(0,1){2}} \put(80,-12){\line(0,1){7}}

\put(10,30){\circle{4}}

\put(80,-10){\circle{4}}\put(10,-10){\circle{4}}

\put(10,-12){\line(0,1){7}}

\put(9,10){\vdots}\put(72,10){$\ddots$}\put(27,10){$\equiv$}

\end{picture}
\vskip-3em\caption{Generalized {\small SWAP} gate; stage 2.
Algorithm description of operation step  on successive
states.}\label{qwerty}
\end{figure}
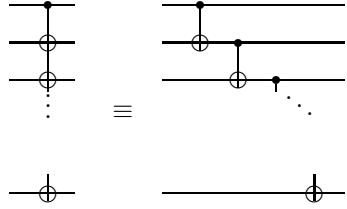

\begin{figure}\setlength{\unitlength}{0.05cm}
\hspace*{2mm}
\begin{picture}(60,80)(-87,-30)
\put(0,20){\line(1,0){37}} \put(38,30){\dots}
\put(38,20){\dots}\put(38,40){\dots}\put(38,-10){\dots}
\put(0,30){\line(1,0){37}} \put(0,-10){\line(1,0){37}}
\put(45,40){\line(1,0){10}} \put(45,30){\line(1,0){10}}
\put(45,20){\line(1,0){10}} \put(45,-10){\line(1,0){10}}
\put(0,40){\line(1,0){37}} \put(10,17){\line(0,1){22}}
\put(10,20){\circle{4}}
\put(50,20){\circle{4}}\put(50,30){\circle{4}}\put(50,-10){\circle{4}}
\put(10,40){\circle*{2}} \put(50,40){\circle*{2}}
\put(50,17){\line(0,1){22}} \put(20,17){\line(0,1){22}}
\put(20,40){\circle*{2}}
\put(10,30){\circle{4}}\put(20,30){\circle{4}}\put(30,30){\circle{4}}\put(20,20){\circle{4}}\put(20,-10){\circle{4}}\put(10,-10){\circle{4}}\put(30,-10){\circle{4}}
\put(30,17){\line(0,1){22}} \put(30,20){\circle{4}}
\put(30,40){\circle*{2}}
\put(10,-12){\line(0,1){5}}\put(20,-12){\line(0,1){5}}\put(30,-12){\line(0,1){5}}\put(50,-12){\line(0,1){5}}
\put(-15,29){$i_1^0$} \put(-15,19){$i_2^0$}
\put(-15,39){$i_0^0$}\put(-15,10){\vdots}\put(19,10){\vdots}\put(29,10){\vdots}\put(9,10){\vdots}\put(49,10){\vdots}
\put(-15,-9){$i_{d-1}^0$}
\end{picture}
\vskip-2em\caption{Generalized {\small SWAP} gate; stage 2, steps
$j = 1,\dots, d-1$.}\label{stage2}
\end{figure}
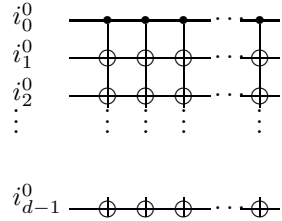

The sequence of values  $i_0^{d-1}, i_1^{d-1}, \dots,
i_{d-1}^{d-1}$ corresponding to the final step of stage 2 are
carried forward as an input sequence for stage 3, step $j = d$.
The algorithm step  keeps the values $i_0^{d-1}, i_1^{d-1}$ and
returns them as outcomes  $i_0^{d}, i_1^{d}$ for step $j = d$. The
remaining systems are then targeted in an iterative process. For
instance, the outcome $i^{d}_2$ for step $j = d$ is given as
$i_0^{d} + i_2^{d-1}$. This value is then stored as the result
$i^{d}_2$ for $e_2$ at stage 3. The outcome state  for
${\mathcal{A}}_0, {\mathcal{A}}_1, {\mathcal{A}}_2$ at stage 3
have thus been determined. To evaluate the result value for
${\mathcal{A}}_3$, the algorithm computes $i_1^{d} + i_3^{d-1}$
and stores this value as the outcome $i^{d}_3$ for  stage 3.
Figure \ref{stage3} illustrates the process that determines the
current state of the algorithm following stage 3, step $j = d$ in
diagrammatic shorthand form for the sequence of generalized
{\small CNOT} gates.

\item[Stage 4]{} $j = d+1$.
\begin{eqnarray}
i^{d+1}_{k} &=& i^{d}_{k}+i_{k+1}^{d}\nonumber\\
i^{d+1}_{d-1} &=& i^{d}_{d-1}; \  k = 0,\dots,d-2.\nonumber
\end{eqnarray}

\begin{figure}\hskip5em\setlength{\unitlength}{0.05cm}
\hspace*{2mm}
\begin{picture}(55,70)(-75,-30)
\put(-14,00){$\equiv$}

\put(-50,20){\line(1,0){27}}
\put(-50,10){\line(1,0){27}}\put(-50,30){\line(1,0){27}}\put(-50,40){\line(1,0){27}}\put(-50,00){\line(1,0){27}}\put(-50,-20){\line(1,0){27}}\put(-50,-30){\line(1,0){27}}

\put(-45,-20){\circle{4}}\put(-45,20){\circle{4}}\put(-45,0){\circle{4}}\put(-32,10){\circle{4}}\put(-32,-30){\circle{4}}
\put(-45,40){\circle*{2}}\put(-32,30){\circle*{2}}\put(-45,-5){\line(0,1){45}}\put(-32,-5){\line(0,1){36}}
\put(-45,-22){\line(0,1){6}}\put(-32,-32){\line(0,1){16}}
\put(-45.6,-10){$\vdots$}\put(-32.6,-10){$\vdots$}

\put(0,20){\line(1,0){47}} \put(48,30){\dots}
\put(48,20){\dots}\put(48,40){\dots}\put(48,10){\dots}\put(48,0){\dots}\put(48,-20){\dots}\put(48,-30){\dots}
\put(0,30){\line(1,0){47}}
\put(0,10){\line(1,0){47}}\put(0,0){\line(1,0){47}}\put(0,-30){\line(1,0){47}}\put(55,-30){\line(1,0){10}}\put(55,0){\line(1,0){10}}
\put(0,-20){\line(1,0){47}} \put(55,40){\line(1,0){10}}
\put(55,30){\line(1,0){10}} \put(55,10){\line(1,0){10}}
\put(47.5,-7){$\ddots$} \put(55,20){\line(1,0){10}}
\put(55,-20){\line(1,0){10}} \put(0,40){\line(1,0){47}}
\put(10,18){\line(0,1){21}}
\put(30,-2){\line(0,1){21}}\put(60,-32){\line(0,1){15}}
\put(10,20){\circle{4}}
\put(20,10){\circle{4}}\put(30,0){\circle{4}}
\put(60,-30){\circle{4}}
\put(10,40){\circle*{2}}\put(40,10){\circle*{2}}\put(40,-4){\line(0,1){14}}
\put(30,20){\circle*{2}}
\put(20,8){\line(0,1){22}} \put(20,30){\circle*{2}}
\put(60,-32){\line(0,1){5}}

\end{picture}
\vskip2em\caption{Generalized {\small SWAP} gate; stage 3, step $j
= d$.}\label{stage3}
\end{figure}
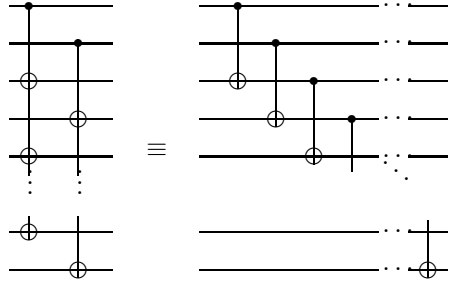

\begin{figure}\setlength{\unitlength}{0.05cm}
\hspace*{3mm}
\begin{picture}(55,60)(-85,-10)
\put(0,20){\line(1,0){37}} \put(38,30){\dots}
\put(38,20){\dots}\put(38,40){\dots}\put(38,10){\dots}\put(38,-10){\dots}
\put(0,30){\line(1,0){37}} \put(0,10){\line(1,0){37}}
\put(0,-10){\line(1,0){37}} \put(45,40){\line(1,0){10}}
\put(45,30){\line(1,0){10}} \put(45,10){\line(1,0){10}}
\put(37.7,3){$\ddots$} \put(45,20){\line(1,0){10}}
\put(45,-10){\line(1,0){10}} \put(0,40){\line(1,0){37}}
\put(20,20){\line(0,1){12}} \put(30,10){\line(0,1){12}}
\put(10,40){\circle{4}}
\put(20,30){\circle{4}} \put(30,20){\circle{4}}
\put(50,-10){\circle*{2}} \put(10,30){\circle*{2}}
\put(30,10){\circle*{2}}
\put(10,30){\line(0,1){12}} \put(20,20){\circle*{2}}
\put(50,-10){\line(0,1){5}}

\end{picture}
\vskip2em\caption{Generalized {\small SWAP} gate; stage 4, step $j
= d+1$.}\label{stage4}
\end{figure}
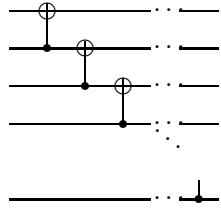

Stage 4 consists of a single step, $j = d+1$,  whose primary
algorithm operation acts as a generalized {\small CNOT} gate on
the $d-1$ consecutive pairs of systems $({\mathcal{A}}_k,
{\mathcal{A}}_{k+1})$ for $k = 0,\dots, d-2$, computing $(i_k^{d}+
i_{k+1}^{d})$ and storing these values as the outcome
$i_{k}^{d+1}$. The value $i_{d-1}^{d}$ is returned as the outcome
$i^{d+1}_{d-1}$.

\item[Stage 5] $j = d+2$.
\begin{eqnarray}
i^{d+2}_{k} &=& i^{d+1}_{k}\nonumber\\
i^{d+2}_{d-1} &=& i^{d+1}_{d-1} +
\sum^{d-2}_{k=0}\eta_{k}i^{d+2}_{k}; \  k = 0,\dots,d-2\nonumber
\end{eqnarray}
  with
\begin{eqnarray}
\sum^{d-2}_{k=0}\eta_{k}i^{d+2}_{k} :=
\sum^{\lfloor\frac{d-2}{2}\rfloor}_{t=0}(d-1)i^{d+2}_{{2t+1}}+\sum^{\frac{d-3}{2}}_{t=0}i^{d+2}_{{2t}}.
\end{eqnarray}

Stage 5 concludes the algorithm with a set of gates targeted on
system ${\mathcal{A}}_{d-1}$ whose current state is represented by
$i_{d-1}^{{d+1}}$. The values $i_0^{{d+2}}, i_1^{{d+2}}, \dots,
i_{d-2}^{{d+2}}$ for the respective  systems ${\mathcal{A}}_0,
{\mathcal{A}}_1, \dots, {\mathcal{A}}_{d-2}$ are unchanged from
their representative values $i_0^{d+1},i_1^{d+1},\dots,
i_{d-2}^{d+1}$ at step $j = d+1$, and are returned as outcomes in
the  final state for step $j = d+2$. The final state of
${\mathcal{A}}_{d-1}$ is given by $i_{d-1}^{{d+2}}=i_{d-1}^{{d+1}}
+ \sum^{d-2}_{k=0}\eta_{k}i^{d+2}_{k} $ $=i_{d-1}^{{d+1}} +
i_0^{{d+1}}+ (d-1)i_1^{{d+1}}+ i_2^{{d+1}}+ (d-1)i_3^{{d+1}}+
\dots + i_{d-3}^{{d+1}}+ (d-1)i_{d-2}^{{d+1}}.$ Thus, for odd
valued $k$ there is a gate with $i^{d+2}_{k}$ as control, and, for
even valued $k$  there are $d-1$ gates with $i^{d+2}_{k}$ as
control. This is represented in figure \ref{stage5}.
\end{description}
 \end{theorem}

\begin{proof} We show that the generalized {\small SWAP} gate algorithm outputs $i^{d+2}_{k} = e_{k+1}$
for $k = 0,\dots,d+2$ and $i^{d+2}_{d-1} = e_0$. At step $j = 0$,
we have it that,
\begin{eqnarray}
{e_{k}} :=  i_{k}^{0};\ k  =  0,...,d-1.
\end{eqnarray}

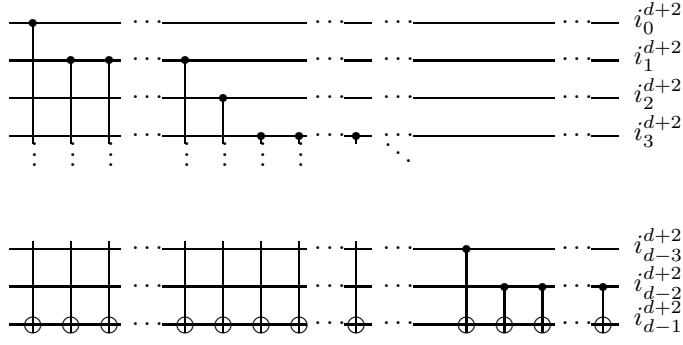
\begin{figure}\hskip3em\setlength{\unitlength}{0.05cm}
\hspace*{2mm}
\begin{picture}(130,95)(190,15)
\put(194,110){\line(1,0){29}} \put(194,100){\line(1,0){29}}
\put(194,90){\line(1,0){29}}
\put(194,50){\line(1,0){29}} \put(194,40){\line(1,0){29}}
\put(194,30){\line(1,0){29}} \put(194,80){\line(1,0){29}}
\put(226,30){\dots} \put(226,40){\dots} \put(226,50){\dots}
\put(226,80){\dots} \put(226,110){\dots} \put(226,100){\dots}
\put(226,90){\dots}
\put(234,110){\line(1,0){38}} \put(234,100){\line(1,0){38}}
\put(234,90){\line(1,0){38}}
\put(234,50){\line(1,0){38}} \put(234,40){\line(1,0){38}}
\put(234,30){\line(1,0){38}} \put(234,80){\line(1,0){38}}
\put(200,28){\line(0,1){24}} \put(210,28){\line(0,1){24}}
\put(220,28){\line(0,1){24}} \put(200,78){\line(0,1){32}}
\put(210,78){\line(0,1){22}} \put(220,78){\line(0,1){22}}
\put(199.5,72){\vdots} \put(209.5,72){\vdots}
\put(219.5,72){\vdots} \put(200,110){\circle*{2}}
\put(200,30){\circle{4}} \put(270,80){\circle*{2}}
\put(270,30){\circle{4}} \put(210,100){\circle*{2}}
\put(210,30){\circle{4}} \put(220,100){\circle*{2}}
\put(220,30){\circle{4}} \put(240,28){\line(0,1){24}}
\put(250,28){\line(0,1){24}} \put(260,28){\line(0,1){24}}
\put(240,78){\line(0,1){22}} \put(250,78){\line(0,1){12}}
\put(260,78){\line(0,1){2}} \put(239.5,72){\vdots}
\put(249.5,72){\vdots} \put(259.5,72){\vdots}
\put(269.5,72){\vdots} \put(240,100){\circle*{2}}
\put(240,30){\circle{4}} \put(250,90){\circle*{2}}
\put(250,30){\circle{4}} \put(260,80){\circle*{2}}
\put(260,30){\circle{4}}
\put(270,28){\line(0,1){24}} \put(270,78){\line(0,1){2}}
\put(285,30){\circle{4}} \put(285,80){\circle*{2}}
\put(285,28){\line(0,1){24}} \put(285,78){\line(0,1){2}}
\put(274,110){\dots} \put(274,100){\dots} \put(274,90){\dots}
\put(274,80){\dots} \put(274,30){\dots} \put(274,40){\dots}
\put(274,50){\dots}
\put(292,72.5){$\ddots$}
\put(282,110){\line(1,0){7}} \put(282,100){\line(1,0){7}}
\put(282,90){\line(1,0){7}}
\put(282,50){\line(1,0){7}} \put(282,40){\line(1,0){7}}
\put(282,30){\line(1,0){7}} \put(282,80){\line(1,0){7}}
\put(292,110){\dots} \put(292,100){\dots} \put(292,90){\dots}
\put(292,80){\dots} \put(292,30){\dots} \put(292,40){\dots}
\put(292,50){\dots}
\put(300,110){\line(1,0){37}} \put(300,100){\line(1,0){37}}
\put(300,90){\line(1,0){37}}
\put(300,50){\line(1,0){37}} \put(300,40){\line(1,0){37}}
\put(300,30){\line(1,0){37}} \put(300,80){\line(1,0){37}}
\put(314,50){\circle*{2}} \put(314,30){\circle{4}}
\put(324,40){\circle*{2}} \put(324,30){\circle{4}}
\put(334,40){\circle*{2}} \put(334,30){\circle{4}}
\put(314,28){\line(0,1){22}} \put(324,28){\line(0,1){12}}
\put(334,28){\line(0,1){12}}
\put(339,110){\dots} \put(339,100){\dots} \put(339,90){\dots}
\put(339,80){\dots} \put(339,30){\dots} \put(339,40){\dots}
\put(339,50){\dots}
\put(347,110){\line(1,0){7}} \put(347,100){\line(1,0){7}}
\put(347,90){\line(1,0){7}}
\put(347,50){\line(1,0){7}} \put(347,40){\line(1,0){7}}
\put(347,30){\line(1,0){7}} \put(347,80){\line(1,0){7}}
\put(350,28){\line(0,1){12}} \put(350,40){\circle*{2}}
\put(350,30){\circle{4}}
\put(358,109){$i_0^{d+2}$} \put(358,99){$i_1^{d+2}$}
\put(358,89){$i_2^{d+2}$} \put(358,79){$i_3^{d+2}$}
\put(358,30){$i_{d-1}^{d+2}$} \put(358,39){$i_{d-2.}^{d+2}$}
\put(358,49){$i_{d-3}^{d+2}$}
\end{picture}
\caption{Generalized {\small SWAP} gate; stage 5, step $j =
d+2$.}\label{stage5}
\end{figure}

At stage 2, step $j = 1$  the algorithm sets $i^1_0 = i^0_0$ and
computes $i^1_1$ as $i^1_1 = i^1_0 + i_1^0$. Similarly, $i^1_2 =
i^1_1 + i_2^0 = i^0_0 + i^0_1 + i^0_2 = \sum_{m=0}^{2}{}i^0_m$.
Therefore, for $k = 1,\dots,d-1$, we have,
\begin{eqnarray}
i_{k}^{1} &=& i_{k-1}^{1} + i^{0}_{k}\nonumber\\
&=& i^{1}_{k-2}+i^{0}_{k-1}+i^{0}_{k}\nonumber\\
&=& i^{1}_{k-3}+i^{0}_{k-2}+i^{0}_{k-1}+i^{0}_{k}\nonumber\\
&=& \dots\nonumber\\
&=& i^{1}_{0}+i^{0}_{1}+i^{0}_{2}+\dots+i^{0}_{k-3}+i^{0}_{k-2}+i^{0}_{k-1}+i^{0}_{k}\nonumber\\
&=& \sum_{m=0}^{k}i_{m}^{0}.
\end{eqnarray}

Next, stage 2 step $j=2$,  repeats the set of gates  of step 1. By
definition $i^{2}_{0} = i^{1}_{0} = i^{0}_{0}$. The case for $k =
1, \dots, d-1$ follows from the algorithm step,
\begin{eqnarray} i^{2}_{k} &=& i^{2}_{k-1} +
i^{1}_{k}  \nonumber \\
&=& i^{2}_{k-2} + i^{1}_{k-1}+i^{1}_{k}\nonumber\\
&=& \dots \nonumber\\
&=& i^{2}_{0}+i^{1}_{1}+i^{1}_{2}+\dots+i^{1}_{k-2} +
i^{1}_{k-1}+i^{1}_{k}\nonumber\\ &=&
i^{0}_{0}+\sum^{1}_{m=0}i^{0}_{m}+\sum^{2}_{m=0}i^{0}_{m} + \dots
+ \sum^{k}_{m=0}i^{0}_{m} \nonumber\\
&=& \sum^{k}_{l=0}\sum^{l}_{m=0}i^{0}_{m}\nonumber\\
&=& \sum^{k}_{m=0}\sum^{k}_{l=m}i^{0}_{m}\nonumber\\
&=& \sum^{k}_{m=0}\sum^{k-m}_{l=0}i^{0}_{m}\nonumber\\
&=& \sum^{k}_{m=0}{k-m+1 \choose 1}i^{0}_{m}.
\end{eqnarray}

For steps $j = 1,\dots,d-1$, we show by induction that $i^{j}_{k}
= \sum^{k}_{m=0}{k-m+j-1\choose j-1}i^{0}_{m}$, $k = 0,\dots,d-1$.
We have shown that this is true for $j=1$. Let $1 \leq j < d-1$
and suppose that
\begin{eqnarray}\label{614}
i^{j}_{k} &=& \sum^{k}_{m=0}{k-m+j-1\choose j-1}i^{0}_{m},
\end{eqnarray}
$k = 0,\dots,d-1$. Now, $i^{j+1}_{0} = i^{j}_{0} = i^0_0$. For $1
\leq k \leq d-1$, we have
\begin{eqnarray}
i^{j+1}_{k} &=& i^{j}_{k} + i^{j+1}_{k-1}\nonumber\\
&=& i^{j}_{k}+ i^{j}_{k-1}+i^{j+1}_{k-2} \nonumber\\
&=& \dots\nonumber\\
&=& i^{j}_{k}+ i^{j}_{k-1}+\dots +i^{j}_{2}+ i^{j}_{1}+ i^{j+1}_{0}\nonumber\\
&=& i^{j}_{k}+ i^{j}_{k-1}+\dots +i^{j}_{2}+ i^{j}_{1}+ i^{j}_{0}.
\end{eqnarray}
Since $i^{j+1}_{0}= i^{j}_{0}$ follows from the algorithm step, we
have it that $ i^{j+1}_{k} = \sum^{k}_{m=0}i^{j}_{m}$. Hence, by
the induction process,
\begin{eqnarray}
i^{j+1}_{k} = \sum^{k}_{m=0}i^{j}_{m} &=& i^0_0 +
\sum^{1}_{l=0}{1-l+j-1\choose
j-1}i^{0}_{l}+\sum^{2}_{l=0}{2-l+j-1\choose
j-1}i^{0}_{l}+\dots\nonumber\\&& +\sum^{k}_ {l=0}{k-l+j-1\choose
j-1}i^{0}_{l}\nonumber\\
&=&\sum^{k}_{m=0}\sum^{m}_{l=0}{m-l+j-1\choose
j-1}i^{0}_{l}\nonumber\\&=&
\sum^{k}_{m=0}\sum^{k}_{l=0}{m-l+j-1\choose
j-1}i^{0}_{l}\nonumber\\
&=& \sum^{k}_{l=0}\sum^{k}_{m=l}{m-l+j-1\choose
j-1}i^{0}_{l}\nonumber\\ &=& \sum^{k}_{l=0}{k-l+j\choose
j}i^{0}_{l}.
\end{eqnarray}
Therefore, the induction step is true for $j+1$,
\begin{eqnarray}
i^{j+1}_{k} &=& \sum^{k}_{m=0}{k-m+j\choose j}i^{0}_{m},
\end{eqnarray}
and the result for stage 2 follows.

The algorithm at stage 3, step $j=d$ yields that
\begin{eqnarray}
i^{d}_{0} &=& i^{d-1}_{0} =   i^{0}_{0}\nonumber\\
i^{d}_{1} &=& i^{d-1}_{1} = \ \sum^{1}_{m=0}{1-m+d-2 \choose
d-2}i^{0}_{m} = (d-1)i_0^0+i_1^0.\end{eqnarray} Implementing the
algorithm step $i^{d}_{k} = i^{d}_{k-2} + i^{d-1}_{k}$ for $k \ =
\ 2,\dots\, d-1$, we have it that
\begin{eqnarray}
i^{d}_{2} &=& i^{d}_{0} + i^{d-1}_{2} = i^{0}_{0} +
\sum^{2}_{m=0}{2-m+d-2 \choose d-2}i^{0}_{m}\nonumber\\
i^{d}_{3} &=& i^{d}_{1} + i^{d-1}_{2} = \ \sum^{1}_{m=0}{1-m+d-2 \choose d-2}i^{0}_{m} + \sum^{3}_{m=0}{3-m+d-2 \choose d-2} i^{0}_{m}\nonumber\\
i^{d}_{4} &=& i^{d}_{2} + i^{d-1}_{4} = \ i^{0}_{0} + \sum^{2}_{m=0}{2-m+d-2 \choose d-2}i^{0}_{m}+ \sum^{4}_{m=0}{4-m+d-2 \choose d-2}i^{0}_{m}\nonumber\\
&\dots& \nonumber
\end{eqnarray}
In particular, for odd valued $k$,
\begin{eqnarray}\label{619}
i^{d}_{k} &=& \sum^{\frac{k-1}{2}}_{t=0}\sum^{2t+1}_{m=0}{2t+1-m +d-2\choose d-2}i^{0}_{m}\nonumber\\
&=& \sum^{\frac{k-1}{2}}_{t=0}\sum^{2t+1}_{m=2t}{2t+1-m +d-2\choose d-2}i^{0}_{m}\ \ \textrm{ (as $d$ is prime)}\nonumber\\
&=& \sum^{\frac{k-1}{2}}_{t=0}{}(d-1)i^{0}_{2t}+i_{2t+1}^{0},
\end{eqnarray}
and, similarly, for even valued  $k$,
 \begin{eqnarray}\label{6110}
i^{d}_{k} &=& \sum^{\frac{k}{2}}_{t=0}\sum^{2t}_{m=0}{2t-m +d-2\choose d-2}i^{0}_{m}\nonumber\\
&=& \sum^{\frac{k}{2}}_{t=0}\sum^{2t}_{m=2t}{2t-m +d-2\choose d-2}i^{0}_{m}\ \ \textrm{ (as $d$ is prime)}\nonumber\\
&=& \sum^{\frac{k}{2}}_{t=0}{}(d-1)i^{0}_{2t-1}+i_{2t}^{0}.
\end{eqnarray}

The next stage, stage 4, step $j = d+1$, of the algorithm is given
by $i^{d+1}_{k}=i^{d}_{k}+i^{d}_{k+1}$ for $k = 0,\dots,d-2$. Let
us consider the value $i^{{d+1}}_{k}$. There are two cases to
note. For even valued $k$, we have it that
\begin{eqnarray}
i^{d}_{k} &=& i^{d}_{k-2}+i^{d-1}_{k}\nonumber\\
&=&
i^{d}_{k-4}+i^{d-1}_{k-2}+i^{d-1}_{k}\nonumber\\
&=& \dots \nonumber\\
&=& \sum^{\frac{k}{2}}_{t=0}{}i^{d-1}_{2t}
\end{eqnarray}
while
\begin{eqnarray}
i^{d}_{k+1} &=&
i^{d}_{k-1}+i^{d-1}_{k+1}\nonumber\end{eqnarray}\begin{eqnarray}
&=& i_{k-3}^{d} + i^{d-1}_{k-1}+i^{d-1}_{k+1}\nonumber\\
&=& \dots \nonumber\\
&=& \sum^{\lfloor\frac{k+1}{2}\rfloor}_{t=0}i^{d-1}_{2t+1}.
\end{eqnarray}
Therefore, $i^{d+1}_{k} = \sum^{k+1}_{t=0}i^{d-1}_{t}$ for even
valued $k$. Correspondingly, for odd valued $k$, $i^{d}_{k} =
\sum^{\frac{k-1}{2}}_{t=0}i^{d-1}_{2t+1}$ while $i^{d}_{k+1} =
\sum^{\frac{k+1}{2}}_{t=0}{}i^{d-1}_{2t}$ and thus $i^{d+1}_{k} =
\sum_{t=0}^{k+1}{i^{d-1}_{t}}$. Hence, we find that
\begin{eqnarray}
i^{d+1}_{k} &=& \sum^{k+1}_{t=0}i^{d-1}_{l}\nonumber\\
&=& \sum^{k+1}_{l=0}\sum^{t}_{m=0}{t-m+d-2 \choose d-2}i^{0}_{m}\nonumber\\
&=& \sum^{k+1}_{m=0}\sum^{k+1}_{l=m}{l-m+d-2\choose d-2}i^{0}_{m}\nonumber\\
&=& \sum^{k+1}_{m=0}{k-m+d \choose d-1}i^{0}_{m}\nonumber\\
&=& i^{0}_{k+1}\ (\textrm{mod} \ d).
\end{eqnarray}
For prime dimensions, $d = p$, recall that under arithmetic modulo
$d$, the coefficients ${k-m+d \choose d-1}$ vanish for $m\ne k+1$.
Therefore,  we  deduce that  $i^{d+1}_{k}=i^{0}_{k+1}$ for $k =
0,\dots,d-2$. When $k = d-1$, we have
\begin{eqnarray}
i^{d+1}_{d-1} &=& i^{d}_{d-1} = \
\sum^{\frac{d-1}{2}}_{t=0}\sum^{2t}_{m=0}{2t-m+d-2 \choose
d-2}i^{0}_m.
\end{eqnarray}

The generalized {\small SWAP} gate algorithm concludes following
stage 5, step $j= d+2$ with the implementation of a sequence of
gates targeted on $i_{d-1}^{d+1}$. For $k = 0,\dots,d-2$, we have
the result
\begin{eqnarray}
i^{d+2}_{k} &=& i^{d+1}_{k}\ =\ i^{0}_{k+1}\ (\textrm{mod}\ d).
\end{eqnarray}
For $k = d-1$, the value  $i^{d+2}_{d-1}$ is given as
\begin{eqnarray}
i^{d+2}_{d-1} &=& i^{d+1}_{d-1} + \sum^{d-2}_{k=0} \eta_{k}i^{d+2}_{k}\nonumber\\
&=&\sum^{\frac{d-1}{2}}_{t=0}\sum^{2t}_{m=0}{2t-m+d-2 \choose
d-2}i^{0}_m+\sum^{d-2}_{k=0}\eta_{k}i^{d+2}_{k}.\nonumber
\end{eqnarray}
To see that this yields the desired result (i.e., $i^{d+2}_{d-1} \
\textrm{mod} \ d  = i^{0}_{0})$, we consider the value
$i^{d+1}_{d-1}\ (\textrm{mod} \ d).$

\begin{lemma}\label{lem}
$i^{d+1}_{d-1}=
\sum^{\frac{d-1}{2}}_{t=0}i^{0}_{2t}+\sum^{\frac{d-1}{2}-1}_{t=0}(d-1)i^{0}_{2t+1}\
(\textrm{mod}\  d) $.
\end{lemma}
\begin{proof} \begin{eqnarray} i^{d+1}_{d-1} &=&
\sum^{\frac{d-1}{2}}_{t=0}\sum^{2t}_{m=0}{2t-m+d-2
\choose d-2}i^{0}_m\nonumber\\
&=& \sum^{d-1}_{m=0}{}\sum^{\frac{d-1}{2}}_{t=\lceil
\frac{m}{2}\rceil}{}{2t-m+d-2 \choose d-2}i^{0}_m.\end{eqnarray}
Since ${2t - m +d-2 \choose d-2} = 0 \ \textrm{mod}\ d$ for $t >
\lceil \frac{m}{2} \rceil$ then
\begin{eqnarray}
i^{d+1}_{d-1} &=&  \sum^{d-1}_{m=0}{}{2\lceil \frac{m}{2}\rceil-m+d-2 \choose d-2}i^{0}_m \nonumber\\
&=&\sum^{\frac{d-1}{2}}_{l=0}i^{0}_{2l}+\sum^{\lfloor\frac{d-2}{2}\rfloor}_{l=0}(d-1)i^{0}_{2l+1}
\ (\textrm{mod}\ d).
\end{eqnarray}
Thus, $i^{d+1}_{d-1} =
\sum^{\frac{d-1}{2}}_{t=0}i^{0}_{2t}+\sum^{\lfloor\frac{d-2}{2}\rfloor}_{t=0}(d-1)i^{0}_{2t+1}
\ (\textrm{mod}\ d). $\end{proof} Finally, by definition of stage
5, we have
$\sum^{d-2}_{k=0}\eta_{k}i^{d+2}_{k} =
\sum^{\lfloor\frac{d-2}{2}\rfloor}_{t=0}(d-1)i^{d+2}_{{2t+1}}+\sum^{\frac{d-3}{2}}_{t=0}i^{d+2}_{{2t}}$.
The value of $i^{d+2}_{d-1}$ is then given by
\begin{eqnarray}
i^{d+2}_{d-1} &=& i^{d+1}_{d-1} + \sum^{d-2}_{k=0}\eta_{k}i^{d+2}_{k}\nonumber\\
&=&
\sum^{\lfloor\frac{d-1}{2}\rfloor}_{t=0}i^{0}_{2t}+\sum^{\lfloor\frac{d-2}{2}\rfloor}_{t=0}(d-1)i^{0}_{2t+1}+
\sum^{\lfloor\frac{d-2}{2}\rfloor}_{t=0}i^{0}_{{2t+1}}+\sum^{\lfloor\frac{d-1}{2}\rfloor}_{t=1}(d-1)i^{0}_{{2t}}.\
\ \ \ \ \ \ \ \ \ \
\end{eqnarray}
Consequently, we have the desired result  $i^{d+2}_{d-1}\
(\textrm{mod}\ d) = i^{d+2}_{0} = i^0_0.$ This completes the proof
of theorem \ref{WilNOT} ensuring that the generalized {\small SWAP
} gate algorithm cyclically permutes the input sequence $i_k^0 =
e_k$, $k = 0,\dots, d-1$ to the output sequence $i^{d+2}_k =
e_{k+1}$, $k = 0,\dots,d-2$ with $i^{d+1}_{d-1} = e_0$.
\end{proof}

We now show that if the generalized {\small SWAP} gate network
swaps an input basis state  then the generalized {\small SWAP}
gate will swap all possible $d^d$ sequences of input states.
\begin{theorem}\label{peter1} Let ${\mathcal{A}}_0,\dots,{\mathcal{A}}_{d-1}$ be
$d$-dimensional systems with bases $\qudit{e_0}_j,\qudit{e_1}_j$,
$\dots,\qudit{e_{d-1}}_j$, $j=0,\dots,d-1$, where
${e_0,\dots,e_{d-1}} \in \Z_d$. Let
${\mathcal{A}}={\mathcal{A}}_0\otimes\dots\otimes
{\mathcal{A}}_{d-1}$. If a network implements a generalized
{\small SWAP} on each basis state $\qudit{a_0a_1\dots a_{d-1}} =
\qudit{a_0}_0\otimes\qudit{a_1}_1\otimes\dots\otimes\qudit{a_{d-1}}_{d-1}$
 of ${\mathcal{A}}$ where $a_0,\dots,a_{d-1}
\in \Z_d$ then the network implements a generalized {\small SWAP}
on any input state $\qudit{\psi} = \qudit{\psi_0}_0
\otimes\qudit{\psi_1}_1\otimes\qudit{\psi_{d-1}}_{d-1}$.
\end{theorem}

\begin{proof} Let $\qudit{\psi_j}_j = \sum_{k_j=0}^{d-1} \alpha_{jk_j}
\qudit{e_{k_j}}_j$, $j=0,\dots,d-1$. Then
\begin{eqnarray}\qudit{\psi}=
\sum_{k_0=0}^{d-1}\dots\sum_{k_{d-1}=0}^{d-1}\alpha_{0k_0}\dots\alpha_{(d-1)k_{d-1}}\qudit{k_0\dots
k_{d-1}}.\end{eqnarray} Now, \begin{eqnarray}
\textrm{SWAP}\qudit{\psi} &=&
\sum_{k_0=0}^{d-1}\dots\sum_{k_{d-1}=0}\alpha_{0k_0}\dots\alpha_{(d-1)k_{d-1}}\textrm{SWAP}
\qudit{k_0\dots k_{d-1}}\nonumber\\&=&
\sum_{k_0=0}^{d-1}\dots\sum_{k_{d-1}=0}^{d-1}\alpha_{0k_0}\dots\alpha_{(d-1)k_{d-1}}\qudit{k_1\dots
k_{d-1}k_0}\nonumber\\&=&
\sum_{k_1=0}^{d-1}\dots\sum_{k_{d-1}=0}^{d-1}\sum_{k_0=0}^{d-1}\alpha_{1k_1}\dots\alpha_{(d-1)k_{d-1}}\alpha_{0k_0}\qudit{k_1\dots
k_{d-1}k_0}\nonumber\\ &=& \qudit{\psi_1}_0
\otimes\dots\otimes\qudit{\psi_{d-1}}_{d-2}
\otimes\qudit{\psi_0}_{d-1}\end{eqnarray} as required. \end{proof}

\begin{figure}\setlength{\unitlength}{0.05cm}
\hspace*{5mm}
\begin{picture}(70,50)(-60,5)




\put(0,40){\line(1,0){90}} \put(0,30){\line(1,0){90}}
\put(0,20){\line(1,0){90}}

\put(20,18){\line(0,1){22}} \put(10,18){\line(0,1){22}}

\put(30,18){\line(0,1){22}}

\put(50,20){\line(0,1){12}} \put(40,30){\line(0,1){12}}


\put(10,30){\circle{4}} \put(10,20){\circle{4}}
\put(10,40){\circle*{2}}

\put(20,40){\circle*{2}} \put(20,20){\circle{4}}
\put(20,30){\circle{4}}

\put(30,40){\circle*{2}} \put(30,20){\circle{4}}

\put(40,30){\circle*{2}} \put(50,20){\circle*{2}}

\put(40,40){\circle{4}} \put(50,30){\circle{4}}

\put(60,40){\circle*{2}} \put(70,30){\circle*{2}}
\put(80,30){\circle*{2}}

\put(60,20){\circle{4}} \put(70,20){\circle{4}}
\put(80,20){\circle{4}}

\put(60,18){\line(0,1){22}} \put(70,18){\line(0,1){12}}
\put(80,18){\line(0,1){12}}





\put(-35,39){$\sum_{i=0}^{2}{a_i\qudit{i}}$}
\put(-35,29){$\sum_{i=0}^{2}{b_i{\qudit{i}}}$}
\put(-35,19){$\sum_{i=0}^{2}{c_i{\qudit{i}}}$}

\put(101,19){$\sum_{i=0}^{2}{a_i\qudit{i}}$}
\put(101,39){$\sum_{i=0}^{2}{b_i{\qudit{i}}}$}
\put(101,29){$\sum_{i=0}^{2}{c_i{\qudit{i}}}$}

\end{picture}
\caption{A {\small SWAP} gate for qutrits. The circuit design
describes the cyclical permutation of three qutrit
states.}\label{poiu}
\end{figure}
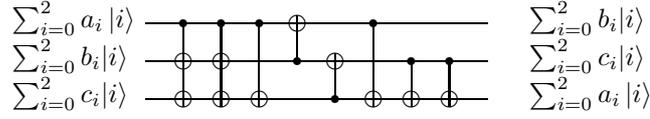

As an example figure \ref{poiu} provides the circuit design of a
generalized {\small SWAP} restricted to qutrits. The quantum
circuit presented comprises of ten two-qutrit {\small CNOT} gates
and represents a concise summary of the main design features of
the generalized {\small SWAP} gate algorithm.

\section{On a generalized   SWAP gate for $d$ other than
prime}\label{wilnotd}

In this section we consider the question of revising the algorithm
construction of \S\ref{WilNOT section} to induce a set of cyclic
permutations of $d$ qudits for $d$ other than prime. However, we
show that such a revision is not possible as it induces an
unavoidable sign change in one subsystem. This question was
motivated by the case $d = 4$ wherein we consider if it was
possible to cyclically permute the states of four $4$-dimensional
subsystem using only instances of the {\small CNOT} gate.


Let us consider a revised quantum circuit algorithm possessing a
stage 1 and stage 2 identical to the generalized swap algorithm of
\S\ref{WilNOT section}. By equation (\ref{614}) (i.e., stage 2 of
generalized
 {\small SWAP} algorithm with step $j=d-1$), the state of the
algorithm is  $i^{d-1}_{0} = i^0_0$, and $i^{d-1}_{k} =
\sum^{k}_{m=0}{}{k-m+d-2 \choose d-2} i^0_m$ for $k =
1,\dots,d-1$. Next, we seek the particular algorithm state output
\begin{eqnarray}\label{even1}
\begin{pmatrix}\hskip-10em{i_{0}^0}\cr \hskip-9.2em (d-1){i_{0}^0} + {i_{1}^0}\cr\hskip-3.35em{i_{0}^0} +
(d-1){i_{1}^0} + {i_{2}^0}\cr\hskip-10em\vdots\cr (d-1){i_{0}^0} + {i_{1}^0} + (d-1){i_{2}^0}+\dots+ {i_{d-1}^0}\cr\end{pmatrix}\label{s3}
\end{eqnarray}
on systems ${\mathcal{A}}_0,\dots,{\mathcal{A}}_{d-1}$,
respectively. Outcome  (\ref{even1}) is generated by the algorithm
process of \S\ref{WilNOT section} for $d$ prime. However,
achieving outcome (\ref{even1}) for $d$ other than prime requires
that stage 3, step $j = d$ of \S\ref{WilNOT section} be revised.
By  revising stage 3, step $j = d$  and taking $i_k^{d-1}$ with
the following linear combination
\begin{eqnarray}
\sum_{s=0}^{k-2}{}a_si^{d-1}_{k-2-s} =
\sum_{s=0}^{k-2}{}\left(a_s\sum^{(k-2)-s}_{m=0}{}{(k-2)-s-m+d-2
\choose d-2}i_m^0\right),
\end{eqnarray}
 where
\begin{eqnarray}
a_s\hskip-.5em &=& \hskip-.52em d-\left[{s+2+d-2 \choose d-2} +
\sum^{s-1}_{t=0}{}a_t{s-t+d-2\choose d-2}\right]+ (-1)^s,
\end{eqnarray}
 we then obtain outcome (\ref{s3}).
\begin{theorem} For d other than prime, the algorithm process at stage 3, step $j = d$ given by
\begin{eqnarray}i_k^{d} &=& i_k^{d-1}+\sum_{s=0}^{k-2}{}a_si_{k-2-s}^{d-1}
\nonumber\\&=& \sum^{k}_{m=0}{}{k-m+d-2 \choose d-2} i^0_m + \sum_{s=0}^{k-2}{}\left(a_s\sum^{(k-2)-s}_{m=0}{}
{(k-2)-s-m+d-2 \choose d-2} i^0_m\right),\nonumber
\end{eqnarray} for $k = 0,\dots,d-1$,
returns outcome (\ref{s3}).
\end{theorem}

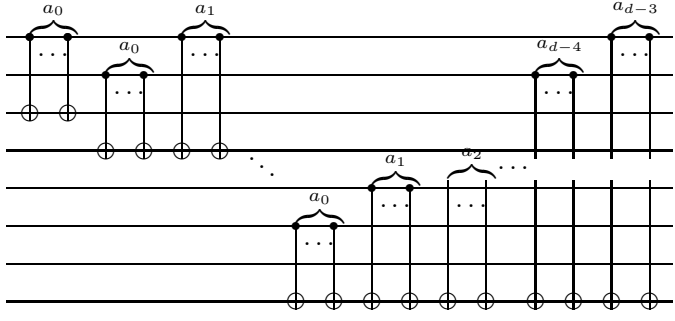
\begin{figure}\hskip5em\setlength{\unitlength}{0.05cm}
\hspace*{2mm}
\begin{picture}(120,110)(200,15)
\put(194,110){\line(1,0){175}} \put(194,100){\line(1,0){175}}
\put(194,90){\line(1,0){175}}
\put(194,50){\line(1,0){175}} \put(194,40){\line(1,0){175}}
\put(194,60){\line(1,0){175}} \put(194,80){\line(1,0){175}}
\put(194,70){\line(1,0){175}} \put(222,95){\dots}
\put(242,105){\dots} \put(202,105){\dots} \put(355,105){\dots}
\put(335,95){\dots}
\put(292,65){\dots}\put(312,65){\dots}\put(272,55){\dots}

\put(234,110){\line(1,0){118}} \put(234,100){\line(1,0){38}}
\put(234,90){\line(1,0){38}}
\put(234,50){\line(1,0){38}}
\put(234,40){\line(1,0){38}}
\put(234,80){\line(1,0){38}}

\put(200,110){$\overbrace{}^{a_0}\begin{matrix} \vspace*{2mm}
\cr\end{matrix}$} \put(220,100){$\overbrace{}^{a_0}\begin{matrix}
\vspace*{2mm} \cr\end{matrix}$}
\put(240,110){$\overbrace{}^{a_1}\begin{matrix} \vspace*{2mm}
\cr\end{matrix}$}

\put(270,60){$\overbrace{}^{a_0}\begin{matrix} \vspace*{2mm}
\cr\end{matrix}$} \put(290,70){$\overbrace{}^{a_1}\begin{matrix}
\vspace*{2mm} \cr\end{matrix}$}
\put(333,100){$\overbrace{}^{a_{d-4}}\begin{matrix} \vspace*{2mm}
\cr\end{matrix}$}
\put(353,110){$\overbrace{}^{a_{d-3}}\begin{matrix} \vspace*{2mm}
\cr\end{matrix}$} \put(310,72){$\overbrace{}^{a_{2}}\begin{matrix}
\vspace*{2mm} \cr\end{matrix}$}

\put(200,110){\circle*{2}} \put(210,110){\circle*{2}}
\put(200,90){\circle{4}} \put(210,90){\circle{4}}
\put(220,80){\circle{4}} \put(230,80){\circle{4}}
\put(250,80){\circle{4}} \put(240,80){\circle{4}}
\put(220,100){\circle*{2}} \put(230,100){\circle*{2}}
\put(250,110){\circle*{2}} \put(240,110){\circle*{2}}

\put(240,78){\line(0,1){32}} \put(250,78){\line(0,1){32}}

\put(230,78){\line(0,1){22}} \put(200,88){\line(0,1){22}}
\put(220,78){\line(0,1){22}} \put(210,88){\line(0,1){22}}

\put(200,110){\circle*{2}} \put(210,110){\circle*{2}}
\put(320,40){\circle{4}} \put(310,40){\circle{4}}
\put(300,40){\circle{4}} \put(290,40){\circle{4}}
\put(280,40){\circle{4}} \put(270,40){\circle{4}}
\put(363,40){\circle{4}} \put(353,40){\circle{4}}
\put(343,40){\circle{4}} \put(333,40){\circle{4}}

\put(300,70){\circle*{2}} \put(290,70){\circle*{2}}
\put(280,60){\circle*{2}} \put(270,60){\circle*{2}}

\put(363,110){\circle*{2}} \put(353,110){\circle*{2}}
\put(343,100){\circle*{2}} \put(333,100){\circle*{2}}

\put(323,75){$\dots$}

\put(320,38){\line(0,1){34}} \put(310,38){\line(0,1){34}}

\put(300,38){\line(0,1){32}} \put(290,38){\line(0,1){32}}
\put(280,38){\line(0,1){22}} \put(270,38){\line(0,1){22}}

\put(333,38){\line(0,1){34}} \put(343,38){\line(0,1){34}}
\put(353,38){\line(0,1){34}} \put(363,38){\line(0,1){34}}

\put(333,38){\line(0,1){34}} \put(343,38){\line(0,1){34}}
\put(353,38){\line(0,1){34}} \put(363,38){\line(0,1){34}}

\put(333,78){\line(0,1){22}} \put(343,78){\line(0,1){22}}
\put(353,78){\line(0,1){32}} \put(363,78){\line(0,1){32}}

\put(257,72.5){$\ddots$}
\put(282,110){\line(1,0){7}} \put(282,100){\line(1,0){7}}
\put(282,90){\line(1,0){7}}
\put(282,50){\line(1,0){7}} \put(282,40){\line(1,0){7}}
\put(282,80){\line(1,0){7}}

\put(300,110){\line(1,0){37}} \put(300,100){\line(1,0){37}}
\put(300,90){\line(1,0){37}}
\put(300,50){\line(1,0){37}} \put(300,40){\line(1,0){37}}
\put(300,80){\line(1,0){37}}

\put(347,110){\line(1,0){7}} \put(347,100){\line(1,0){7}}
\put(347,90){\line(1,0){7}}

\put(347,50){\line(1,0){7}}

\put(347,80){\line(1,0){7}}
\end{picture}
\vskip-1em\caption{Stage 3, step $j=d$. Circuit description
representing the modified stage of the generalized {\small SWAP}
gate for $d$ other than prime. }\label{evenstage3}
\end{figure}

\begin{proof} For $k = 0,1$, we have it that $i^{d}_{0} =
i^{d-1}_{0}$ and $i^{d}_{1} = i^{d-1}_{1}$. Thus, the states $e_0$
and $e_1$ are given as  $i^{0}_{0}$ and $(d-1)i^{0}_{0}+
i^{0}_{1}$ respectively. The state $e_2$ is written as
\begin{eqnarray} i^{d}_{2} &=& \sum_{m=0}^{2}{}{2-m+d-2 \choose d-2} i^0_m + a_0
i^0_0\nonumber\\
&=& \left( {d\choose d-2} + (d - {d\choose d-2} +1)\right)i^0_0 + (d-1)i_1^0 + i_2^0\nonumber\\
&=& i_0^0 + (d-1)i^{0}_{1}+ i^{0}_{2}\  (\textrm{mod}\ d).
\end{eqnarray}
We show by induction that, for $k = 0\ (\textrm{mod}\ 2)$,
\begin{eqnarray}i_k^{d} &=& \sum^{k}_{m=0}{}{k-m+d-2 \choose d-2} i^0_m +
\sum_{s=0}^{k-2}{}\left(a_s\sum^{(k-2)-s}_{m=0}{}{(k-2)-s-m+d-2
\choose d-2} i^0_m\right)\nonumber\\ \hskip2em&=& i^0
_0+(d-1)i^0_1+i^0_2+\dots+(d-1)i^0_{k-1}+i^0_k \ (\textrm{mod} \
d)
\end{eqnarray} and for $k \ne 0\ (\textrm{mod}\ 2)$,
\begin{eqnarray}i_k^{d} &=&
(d-1)i^0_0+i^0_1+(d-1)i^0_2+\dots+(d-1)i^0_{k-1}+i^0_k\
(\textrm{mod}\  d).\end{eqnarray}
We have shown that this is true for $k=0,1,2$. Suppose $0 \leq k
\leq d-2$ and further suppose that
\begin{eqnarray}
i_k^{d} &=& \sum^{k}_{m=0}{}{k-m+d-2 \choose d-2} i^0_m + \sum_{s=0}^{k-2}{}\left(a_s\sum^{(k-2)-s}_{m=0}{}{(k-2)-s-m+d-2 \choose d-2} i^0_m\right)\nonumber\\
&=& \sum_{m=0}^{k}{}(-1)^{k-m}i^0_m\nonumber\\
&=& 
i_0^0 + (d-1)i^{0}_{1}+ i^{0}_{2}+\dots+ (d-1)i^{0}_{k-1}+
i^{0}_{k}\  (\textrm{mod}\ d)\end{eqnarray} {for} $k = 0 \
(\textrm{mod}\ 2)$, {and} \begin{eqnarray} i_k^{d} &=&(d-1)i_0^0 +
i^{0}_{1}+ (d-1)i^{0}_{2}+\dots+ (d-1)i^{0}_{k-1}+ i^{0}_{k}\
(\textrm{mod}\ d)
\end{eqnarray} $\textrm{for}\ k \ne 0\ (\textrm{mod}\ 2)$.
Therefore, for $j = d$, we have,
\begin{eqnarray}
i^{d}_{k+1}{}&=& \sum^{k+1}_{m=0}{}{k+1-m+d-2\choose d-2}i^0_{m}+
\sum_{s=0}^{k-1}{}\left
(a_s\sum^{(k-1)-s}_{m=0}{}{(k-1)-s-m+d-2\choose
d-2}i^0_{m}\right)\nonumber\\&=&
\sum^{k}_{m=0}{}\left({k-m+d-2\choose d-2}i^0_{m+1}+{k+1+d-2\choose d-2}i^0_{0}\right)\nonumber\\
&& + \sum_{s=0}^{k-1}{}a_s\left(\left(\sum^{(k-2)-s}_{m=0}{}{(k-2)-s-m+d-2\choose d-2}i^0_{m+1}\right)+{(k-1)-s+d-2\choose d-2}i^0_{0}\right)\nonumber\\
&=& \sum^{k}_{m=0}{}(-1)^{k-m}i_{m+1}^0 + \left({k+1+d-2\choose
d-2}+\sum^{k-1}_{s=0}{}a_s{k-1-s+d-2\choose d-2}\right)i^{0}_{0}.
\end{eqnarray}
Recall that the binomial coefficients of $i_k^{d-1} =
\sum^{k}_{m=0}{}{k-m+d-2 \choose d-2} i^0_m$ are precisely those
coefficients of $i^{d-1}_{k+1} = \sum_{m=0}^{k+1}{}{k+1-m+d-2
\choose d-2} i^0_m$ for $m = 1,\dots,k+1$. Hence, the particular
combination of systems ${\mathcal{A}}_{(k-2)-s}$ that return the
state $i^{d}_{k}= i_0^0 + (d-1)i^{0}_{1}+ i^{0}_{2}+\dots+
(d-1)i^{0}_{k-1}+ i^{0}_{k}\ (\textrm{mod}\  d)$ is the
combination that yields the a similar sequence on $i^{d}_{k+1}$
for $m=1,\dots,k+1$. With $k = 0 \ (\textrm{mod}\ 2)$, then for
$i^{d}_{k+1}$ we require that the scalar value for $i_0^0$
degenerates to $d-1$ (mod $d$). Thus, for $m=0$ and by definition
of $a_s$, we have
\begin{eqnarray}\hskip-5em&& \left({(k+1)+d-2\choose d-2} + \sum_{s=0}^{k-1}{}a_s{(k-1)-s+d-2\choose d-2}\right)i^0_0\nonumber\\
&&= \left({(k+1)+d-2\choose d-2} + \sum_{s=0}^{k-2}{}\left(a_s{(k-1)-s+d-2\choose d-2}\right)+a_{k-1}\right)i^0_0\nonumber\\
&&= \Biggl({(k+1)+d-2\choose d-2} + \sum_{s=0}^{k-2}{}a_s{(k-1)-s+d-2\choose d-2} \nonumber\\
&&\hskip1em+\left(d - \left[{(k+1)+d-2\choose d-2} + \sum_{s=0}^{k-2}{}a_s{(k-1)-s+d-2\choose d-2}\right] + (-1)^{k+1}\right)\Biggr)i^0_0\nonumber\\
&&= (-1)^{k+1} i^0_0 \ (\textrm{mod}\ d).
\end{eqnarray}
Hence, $i^{d}_{k+1}=\sum_{m=0}^{k+1}{}(-1)^{k+1-m}i^0_{m}$
(\textrm{mod}\ $d$), and the result follows.
\end{proof}
We continue with stage 4 of the generalized {\small SWAP} gate
algorithm of \S\ref{WilNOT section};
\begin{eqnarray}
i^{d+1}_{k}&=&i^{d}_{k}+i^{d}_{k+1}\end{eqnarray}
 for $k = 0,\dots,d-2$ and
\begin{eqnarray}
i^{d+1}_{d-1}&=&i^{d}_{d-1}+\sum_{m=0}^{d-2}{(-1)^{d-1-s}i^{0}_{m}}.
\end{eqnarray} Finally, revising  stage 5 of \S\ref{WilNOT section} so that
$\sum^{d-1}_{k=1}\eta^{*}_{k}i^{d+2}_{k} =
\sum^{\lfloor\frac{d-1}{2}\rfloor}_{t=0}
(d-1)i^{0}_{2t+1}+\sum^{\frac{d-2}{2}}_{t=0} i^{0}_{2t}$, we
obtain the state
\begin{eqnarray}\label{even3}\left(
  {i_{1}^0},{i_{2}^0},{i_{3}^0},  \dots, {i_{d-1}^0},  (d-1){i_{0}^0}
\right).\end{eqnarray} Unfortunately, we have not achieved a
generalized {\small SWAP} for $d$ other than prime as the revised
algorithm has produced a  state that includes a sign change (i.e.,
$-1$ (mod $d$)) in subsystem ${\mathcal{A}}_{d-1}$. The following
argument  shows that a different algorithm would be required. On
obtaining the outcome (\ref{even3}), for $d$ other than prime,  no
sequence of generalized {\small{CNOT}} gates will return  the
desired cyclic permutation. To show this claim, consider the more
general case of outcome (\ref{even3})  given by  the revised
algorithm;
\begin{eqnarray}\label{even4}\left(
\xi{i_{1}^0}, \xi{i_{2}^0}, \xi{i_{3}^0}, \dots,\xi{i_{d-1}^0},
(d-\xi){i_{0}^0} \right).\end{eqnarray} Consider the pairs
$(\xi{i^0_{k}},\xi{i^0_{k+1}})$ for  $k \in \{1,\dots,d-3\}$
together with the final  pair $(\xi{i^0_{d-1}},
(d-\xi)\xi{i^0_{0}})$. Given $(\xi{i_{k}^0},\xi{i_{k+1}^0})$ for
$k \in \{1,\dots,d-3\}$, and a {\small CNOT} mapping
 that targets  $e_{k+1}$, we have it that
$(\xi{i_{k}^0},\xi{i_{k+1}^0})$ $ \mapsto
(\xi{i_{k}^0},\xi{i_{k}^0}+\xi{i_{k+1}^0}).
$
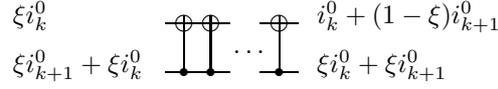
\begin{figure}
\setlength{\unitlength}{0.08cm} \hspace*{35mm}
\hskip-4.5em\vskip3em \setlength{\unitlength}{0.05cm}
\hspace*{63mm}
\begin{picture}(20,20)(20,10)
\put(-40,35){$\xi{i^0_{k}}$}
\put(-40,22){$\xi{i^0_{k+1}}+\xi{i_{k}^0}$}
\put(40,35){${i_{k}^0}+(1-\xi){i_{k+1}^0}$}
\put(40,22){$\xi{i_{k}^0}+\xi{i_{k+1}^0}$}
\put(0,22){\line(1,0){17}} \put(0,35){\line(1,0){17}}
\put(25,22){\line(1,0){10}} \put(25,35){\line(1,0){10}}

\put(5,21){\line(0,1){16}} \put(5,22){\circle*{2}}
\put(5,35){\circle{4}} \put(12,21){\line(0,1){16}}
\put(12,22){\circle*{2}} \put(12,35){\circle{4}}
\put(30,21){\line(0,1){16}} \put(30,22){\circle*{2}}
\put(30,35){\circle{4}} \put(18,27){\dots}
\end{picture}
\caption{$P_{\xi}-1$ generalized {\small CNOT} gates on pairs
$(e_k, e_{k+1})$, $k \in \{0,\dots,d-3\}$}\label{aaaa}
\end{figure}
Denote by $P_{\xi}$ the inverse  of $\xi$ (mod $d$), whence,
$P_{\xi}\xi =$ 1 (mod  $d$). Applying  $P_{\xi} - 1$ gates, see
figure \ref{aaaa}, to target $\xi{i_{k}^0}$ in each pair yields
\begin{eqnarray}
(\xi{i_{k}^0}, \xi{i_{k}^0}+\xi{i_{k+1}^0}) &\mapsto& (P_{\xi}\xi{i_{k}^0}+(P_{\xi}-1)\xi{i_{k+1}^0}, \xi{i_{k}^0}+\xi{i_{k+1}^0})\nonumber\\
&=&({i_{k}^0}+(1-\xi){i_{k+1}^0},
\xi{i_{k}^0}+\xi{i_{k+1}^0}).\label{gates1} \end{eqnarray} In
eliminating  $\xi{i_{k}^0}$ in equation (\ref{gates1}), we apply
$d-\xi$ gates that target $\xi{i_{k}^0}+\xi{i_{k+1}^0}$;
\begin{eqnarray}
({i_{k}^0}+(1-\xi){i_{k+1}^0},\xi{i_{k}^0}+\xi{i_{k+1}^0})&\mapsto&
 ({i_{k}^0}+(1-\xi){i_{k+1}^0}, \xi{i_{k}^0}+\xi{i_{k+1}^0} + (d-\xi)({i_{k}^0}+\nonumber\\&& (1-\xi){i_{k+1}^0})\nonumber\\
 &=&({i_{k}^0}+(1-\xi){i_{k+1}^0}, \xi{i_{k+1}^0} + (d-\xi)(1-\xi){i_{k+1}^0})\nonumber\\
&=& ({i_{k}^0}+(1-\xi){i_{k+1}^0},\xi{i_{k+1}^0}  +
(-\xi+\xi^{2}){i_{k+1}^0})\nonumber\\ &=&
({i_{k}^0}+(1-\xi){i_{k+1}^0},\xi^{2}{i_{k+1}^0}).\label{gates2}
\end{eqnarray}
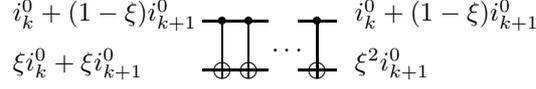
\begin{figure}
\setlength{\unitlength}{0.08cm} \hspace*{35mm}
\hskip-4.5em\vskip3em \setlength{\unitlength}{0.05cm}
\hspace*{63mm}
\begin{picture}(20,20)(10,10)
\put(-50,35){${i_{k}^0}+(1-\xi){i_{k+1}^0}$}
\put(-50,22){$\xi{i_{k}^0}+\xi{i_{k+1}^0}$}
\put(40,35){${i_{k}^0}+(1-\xi){i_{k+1}^0}$}
\put(40,22){$\xi^{2}{i_{k+1}^0}$} \put(0,22){\line(1,0){17}}
\put(0,35){\line(1,0){17}} \put(25,22){\line(1,0){10}}
\put(25,35){\line(1,0){10}}

\put(5,20){\line(0,1){16}} \put(5,22){\circle{4}}
\put(5,35){\circle*{2}} \put(12,20){\line(0,1){16}}
\put(12,22){\circle{4}} \put(12,35){\circle*{2}}
\put(30,20){\line(0,1){16}} \put(30,22){\circle{4}}
\put(30,35){\circle*{2}} \put(18,27){\dots}
\end{picture}
\caption{$d-\xi$ generalized {\small CNOT} gates  on on pairs
$(e_k, e_{k+1})$, $k \in \{0,\dots,d-3\}$.}\label{aaa}
\end{figure}
Similarly, applying the set gates as outlined in figures
\ref{aaaa} and \ref{aaa} to the pair $(\xi{i^0_{d-1}},
(d-\xi){i^0_{0}})$, we obtain $(i^0_{d-1}+(\xi-1)i_0^0,
-\xi^2i_0^0).$ Thus, we have
\begin{eqnarray}\label{even6}&\hskip-16em\left(
\xi{i_{1}^0},   \xi{i_{2}^0},  \dots,  \xi{i_{d-3}^0},
\xi{i_{d-2}^0},  \xi{i_{d-1}^0},  (d-\xi){i_{0}^0}
\right) \mapsto& \nonumber\\
&\hskip-2em\left({i_{1}^0}+(1-\xi){i_{2}^0},
\xi^2{i_{2}^0},\dots,{i_{d-3}^0}+(1-\xi){i_{d-2}^0},
\xi^2{i_{d-2}^0}, i^0_{d-1}+(\xi-1)i_0^0, -\xi^2{i_{0}^0}
\right).\ & \ \end{eqnarray} Since the scalar values $\xi^{2}$ and
$-\xi^{2}$  can not  be both 1 (mod $d$), it seems that any
mapping will fail to return a state with scalars all equal to
unity. This might suggest that a generalized {\small SWAP} gate
composed entirely in terms of generalized {\small CNOT} gates
 may not be possible for $d$ other than prime.

\section{Conclusion}

We discussed the construction of a generalized {\small SWAP} gate
that cyclically permutes the states of $d$ qudit subsystems for
$d$ prime. The design restricted itself to only using instances of
the generalized {\small CNOT} gate, and the analysis made great
use of modular binomial relationships. Lastly, we illustrated how
the generalized {\small SWAP} gate design may be revised to yield
certain permutations  of $d$ qudits for $d$ other than prime.

\begin{acknowledgements}
It is a pleasure to acknowledge assistance received from Prof.
Peter Wild. The author would also like to thank Prof. Matthew G.
Parker and Prof. R\"udiger Schack for helpful comments and
suggestions. This work was completed while at the Dept. of
Mathematics, Royal Holloway, University of London.
\end{acknowledgements}

\label{lastpage}
\end{document}